\documentclass{ws-mpla}

\begin{document}
\def\A{{\cal A}}\def\L{{\cal L}}\def\O{{\cal O}}
\def\la{\langle}\def\ra{\rangle}
\def\be{\begin{eqnarray}}\def\ee{\end{eqnarray}}
\def\lsim{\mathrel{\rlap{\lower3pt\hbox{\hskip1pt$\sim$}}
     \raise1pt\hbox{$<$}}} %less than or approx. symbol
\def\gsim{\mathrel{\rlap{\lower3pt\hbox{\hskip1pt$\sim$}}
     \raise1pt\hbox{$>$}}} %greater than or approx. symbol

\markboth{Won-Gi Paeng and Mannque Rho}
{Toward an EFT for cold CBM}

%%%%%%%%%%%%%%%%%%%%% Publisher's Area please ignore %%%%%%%%%%%%%%

%%%%%%%%%%%%%%%%%%%%%%%%%%%%%%%%%%%%%%%%%%%%%%%%%%%%%%%%%%%%%%%%%%%

\title{TOWARD AN EFFECTIVE FIELD THEORY\\ FOR COLD COMPRESSED BARYONIC MATTER \footnote{The World Class University (WCU) Lecture given in the Fall 2009.}
}

\author{\footnotesize WON-GI PAENG
%\footnote{}
}

\address{Department of Physics, Hanyang University\\
133-791 Seoul, Korea\\ wgpaeng0@gmail.com}

\author{MANNQUE RHO}

\address{Institut de Physique Th\'eorique,  CEA Saclay\\
91191 Gif-sur-Yvette C\'edex, France\\
 and Department of Physics, Hanyang University\\ 133-791 Seoul, Korea\\
mannque.rho@cea.fr}

\maketitle

\begin{abstract}
This is an extended version of the note taken by the first author (W.-G.P.) on a lecture given  by the second author (M.R.) as a first part of the series on ``Hadronic Matter Under Extreme Conditions," the principal theme of the WCU-Hanyang Program. It covers the attempts to go in a framework anchored on effective field theory of QCD from zero density to the nuclear matter density and slightly beyond, with the ultimate goal of arriving at the density relevant to compact stars, including chiral phase transition and quark matter. The focus is on the conceptual aspects rather than detailed ``fitting"  of the data on the kinds of physics that are being addressed to in radioactive-ion-beam machines in operation as well as in project (such as `KoRIA' in Korea) and will be explored at such forthcoming accelerators as FAIR/GSI. The approach presented here is basically different from the standard ones found in the literature in that the notion of hidden local symmetry -- which underlies the chiral symmetry of the strong interactions -- and its generalization to dual gravity description involving infinite tower of hidden gauge fields are closely relied on.

\keywords{Cold dense matter; chiral symemtry; BR scaling; Landau Fermi liquid; double decimation; symmetry energy; FAIR; KoRIA}
\end{abstract}

\ccode{PACS Nos.:  21.65.+f, 71.10.Ay, 11.30.Rd, 21.65.Qr, 21.30.Fe, 13.75.Jz, 97.60.Jd
}

\section{Introduction}

What we would like to do in this lecture is to figure out what happens in dense matter at zero temperature expected in compact stars. This lecture is about what we understand up to the present time, and what we would like to understand in the future.

We start with a cartoon panorama.
%Figure 1_phase diagram%
\begin{figure}[ht]
\centering
\includegraphics[angle=0, height=0.4\textwidth]{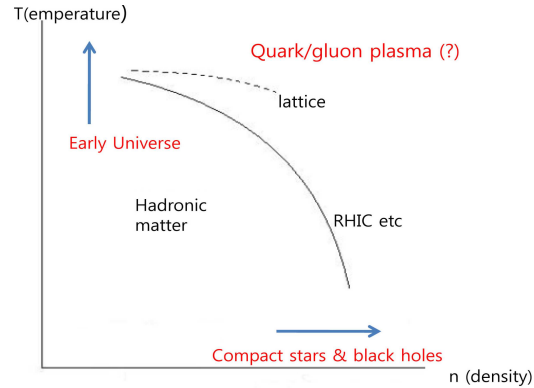}
\caption[0]{The phase diagram of hadronic matter in temperature and density that is being studied.
The solid line describes experimental effort, e.g., SPS/CERN, RHIC and the dotted line lattice QCD.}\label{figure1}
\end{figure}

  We are often shown, very roughly,  a phase diagram like Figure 1, which is given in terms of density vs. temperature. When you go along the temperature axis (and at zero density), you are probing the Early Universe. Along the temperature axis at near zero density, nothing drastic happens. There is no real phase transition of the kind naively predicted and widely heralded by theorists in the past. This feature is confirmed by lattice QCD calculation.  Experiments are probing a wide range of temperature {\em and} density (indicated by solid line in Figure 1). We don't know actually where what one would call genuine ``quark gluon plasma" (QGP) is. But one expects that QGP should be there somewhere in the figure. In a phase where there are weakly or non-interacting quarks and gluons with asymptotic freedom, you can calculate things in perturbation theory. So thinking that one will get quickly to QGP, people have done such calculations, a huge number of them.

 {Many people were dead wrong}. The naive QGP is not visible yet. It's a lot more subtle. Whatever it is, it is strongly coupled up to quite high temperature and does not even require QCD proper to describe it. {Some sort of universality involving conformality is in action, blissfully ignoring whether certain symmetries such as supersymmetry are there or not. This seems to mean that one is not really learning anything significant here.}

Furthermore, whatever physics it is that you are describing in terms of quarks and gluons of QCD, at low energy regime, you can also describe in terms of hadrons and other degrees of freedoms which are actually measured in experiments. Lots of people are calculating what happens in low-density non-zero temperature regime which are being and will be explored by experiments, at e.g., SPS/CERN, RHIC, ALICE/LHC etc. Here lattice plays a crucial role in guiding theorists.

But lattice has tremendous difficulty handling density and in particular, the high density appropriate for compact stars seems out of reach of lattice techniques. There are no experiments to guide theorists beyond the normal nuclear matter density $n_0$ -- and the experimental facilities for this regime are yet to come, so there are no trustful model approaches either.  We did not put in the figure anything {\em novel} dealing with density. You see why. It is a fair picture of the current status. There is nothing much known there. Dense matter is a totally virgin field to explore and as indicated on Fig.~2, this is the WCU objective. The only thing we are sure of is the nuclear matter at $n=n_0=0.16$ fm$^{-3}$. There is also a large effort at various laboratories in the world, e.g., GANIL and RIB machines (e.g., also forthcoming KoRIA?), to unraveling the gas-liquid phase transition at lower density.

\section{Phenomenology with Increasing Density}
%Figure 2_KORIA Experiment%
\begin{figure}[hb]
\centering
\includegraphics[angle=0, height=0.3\textwidth]{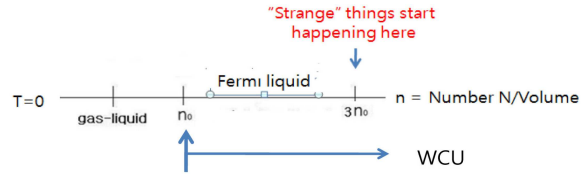}
\caption[0]{First there is the gas-liquid phase transition point, then the (Fermi) liquid phase. Nuclear physics up to the nuclear matter density $n_0 = 0.16 {\rm fm}^{-3}$ is {\em very well} established. From $n_0$ up to the chiral restoration density $n_\chi$ is the density domain for the WCU-Hanyang Program.}\label{figure2}
\end{figure}
\subsection{Nuclear physics up to $\sim 3 n_0$}
Without much ado, let's jump into a specific case raised by Bao-An Li's talks given here. This case illustrates all the interesting features we are particularly interested in. Figure 3A shows the behavior of nuclear symmetry energy -- which figures importantly in neutron-rich nuclei, compact stars etc. -- at density above $n_0$ in various different models~\cite{xiao},\cite{thorsson},\cite{Das}). It shows how widely diverging are the predictions for the symmetry energy given by various models that all fit nuclear matter properties up to, and perhaps slightly above, nuclear matter density. The divergence simply shows how little constrainable the symmetry energy is both by theory and by experiment at higher density. The situation is not so bad for the energy for symmetric ($N=Z$) part (although beyond $n_0$ there is certain spread in prediction), so we will focus on the symmetry energy that enters where there is neutron excess which is the tendency in Nature, both in laboratories and in compact stars.

%Figure 3 model enhancement%
\begin{figure}[htb]
\centering
\includegraphics[angle=0, height=0.25\textheight]{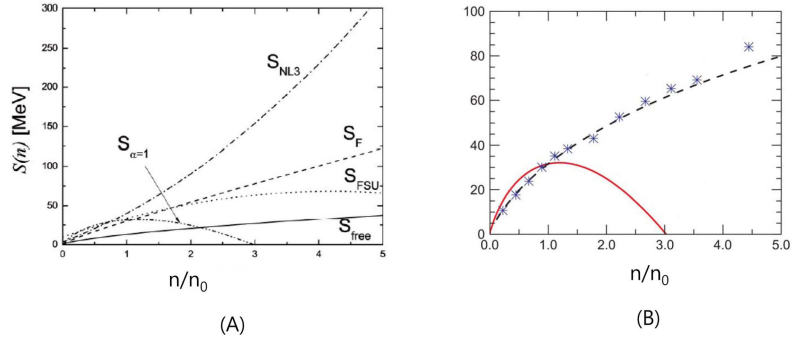}
\caption[0]{Diverse symmetry energies predicted by different models (A) which are caricatured by the simplified one (B).\cite{Kyungmin}.}\label{figure3}
\end{figure}

In most of models in the literature, the symmetry energy increases steadily with increasing density. But the one that fits the available $\pi^-/\pi^+$ ratio measured by FOPI/GSI and analyzed by Bao-An Li~\cite{Das} --  there is only one data at the moment -- differs widely from the rest (call it ``canonical") in that it increases up to certain density, say, around $\sim 1.3n_0$ and then turns over and decreases steadily, going to zero at $n\approx 3n_0$. Let us call it ``supersoft" following Bao-An Li. For convenience, we just replace this complicated picture by the simplified one  Fig.~3B. We will address the problem referring to Figure 3B.

Suppose Li's symmetry energy $S_{\alpha=1}$ (called ``supersoft") is what's given by Nature. There are then several serious consequences:
\begin{enumerate}
\item With the standard description of kaon-nuclear interactions~\cite{BLRstar} -- largely based on chiral perturbation theory -- which predicts that an anti-kaon mass drops continuously in compact star matter as density increases, there cannot be kaon condensation at any reasonably low enough density to be relevant to neutron-star physics.
\item Unless one invokes non-Newtonian gravity, there would be no stable neutron stars of the range of masses and radii observed so far, clearly at odds with nature\cite{Wen}.
\item Then all the ``exciting things" like the Bethe-Brown maximum mass of neutron stars, the maximization of black holes in the Universe and the cosmological natural selection discussed in \cite{BLR-PRL} will be nothing but  red herrings.
\end{enumerate}
These issues will pose great challenges to nuclear physicists. The ``standard" chiral perturbation approach typically gives the kaon mass behaving like what's shown as ``$\chi PT$" in Figure 4. The kaon with that sort of dropping mass cannot possibly condense in compact-star matter, since the electron chemical potential $\mu_e$ is not high enough to make the ``electron decay" $e\rightarrow K^- +\nu$. Such a ``supersoft" symmetry energy will typically make a havoc to neutron stars as one can verify by sticking it into the TOV equation. As Li suggests, one can make drastic measures to ``save" the soft symmetry energy. One way-out suggested by Li is to introduce non-Newtonian gravity without disturbing what's established with standard gravity theory. This spells like a disaster but it could also be an exciting opportunity for the future.\footnote{Indeed, it is a fashionable idea that Newtonian gravity arises as an ``emergent" phenomenon through a holographic scenario. This possibility offers an avenue for a conceptual breakthrough for string theory\cite{verlinde} as well as for loop quantum gravity theory\cite{smolin}.} Because when something goes wrong, there is a ground for a breakthrough: One now has to find out whether the $S_{\alpha=1}$ is what Nature actually says. Experimentally, a part of the symmetry energy -- the lower density part -- could be measured at radioactive ion beam (RIB) machines in construction all over the world, including the one Koreans are going to build, called ``KoRIA," and with precision measurements, one could check whether the FOPI experiment is actually constraining the symmetry energy that drastically. There are also big theoretical issues, the solutions of which could provide ways out of the disaster. We will discuss below two related possibilities based on BR scaling\cite{br91} and three-body (and many-body) forces that would solve the problem even if the supersoft symmetry were {\em correct up to the measured density}.

%Figure 4_kaon mass%
\begin{figure}[ht]
\centering
\includegraphics[angle=0, width=0.5\textwidth]{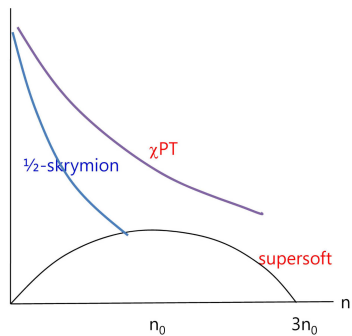}
\caption[0]{The electron chemical potential $\mu_e$  which is directly related to the nuclear symmetry energy corresponding to ``supersoft," the ``standard" kaon mass labeled $\chi PT$ predicted by the class of models belonging to chiral perturbation theory and the kaon mass predicted in dense half-skyrmion matter. The ``standard" kaon mass does not meet with the chemical potential given by ``supersoft," so there will no kaon condensation. However the half-skyrmion phase makes the kaon mass drop sufficiently fast that it could trigger kaon condensation.}\label{figure4}
\end{figure}

Suppose kaons do condense in compact stars at a density near $3n_0$. Below we will discuss how this can happen because of some highly non-perturbative mechanism recently discovered. If this happens, it has an interesting implication on laboratory experiments.

Kaon condensation can take place because there is effectively strong kaon-nuclear attraction. Toshi Yamazaki, who is a well-known Japanese nuclear experimentalist whom the lecturer (MR) admires (one of those rare nuclear experimentalists who do not take seriously what theorists say),  predicted that if the anti-kaon nuclear potential were as deep as
\begin{eqnarray}
V_{KN} (n_0) \lsim -200 {\rm MeV}
\end{eqnarray}
then there could be highly bound kaon-nuclear systems with an average density much higher than that of nuclear matter. His argument\cite{Yamazaki}, as is being debated, is not a ``kosher," as some of the chiral perturbation theorists who do extensive chiral perturbation calculations with various approximations do not agree with his prediction. But one should recognize that chiral perturbation theory, which is a QCD for certain processes (involving pseudo-Goldstone boson) in some low-energy kinematic domain\cite{Leutwyler}, may not be able to access high density regime. In fact, one doubts that what comes out of chiral perturbation theory at low order and involving a limited number of degrees of freedom could be sufficiently reliable. As we will discuss below, there can be certain nonperturbative mechanisms that cannot be accessed by a standard chiral perturbation technique, namely, a phase change of the baryonic matter into a half-skyrmion or half-instanton matter with chiral symmetry ``restored" in the sense that the quark condensate vanishes but with  pions fluctuating on top of the modified ground state (or ``vacuum") and with quarks/gluons still confined in hadrons, contrary to the standard scenario. In such a phase, kaons can become very light and could condense at low density.

\subsection{Half-skyrmions and nuclear physics at $\sim 3n_0$}
As was explained in previous lectures, in the absence of first-principle calculations anchored on QCD in approaching dense matter, the best one can do is to start with hidden local symmetry Lagrangian\cite{harada}. It would be much more realistic to do it with the 5D Yang-Mills action supplemented with the Chern-Simons term, either deconstructed from low-energy or reduced from string theory from high energy. We have had a number of lectures on this latter matter in this WCU series. Unfortunately we cannot yet handle the 5D theory for technical reasons we have heard quite a bit about, such as $1/N_c$, $1/\lambda$ corrections etc., so let us do the 4D theory, i.e., the Harada-Yamawaki (HY) HLS theory.
%Figure 5%
\begin{figure}[ht]
\centering
\includegraphics[angle=0, width=0.8\textwidth]{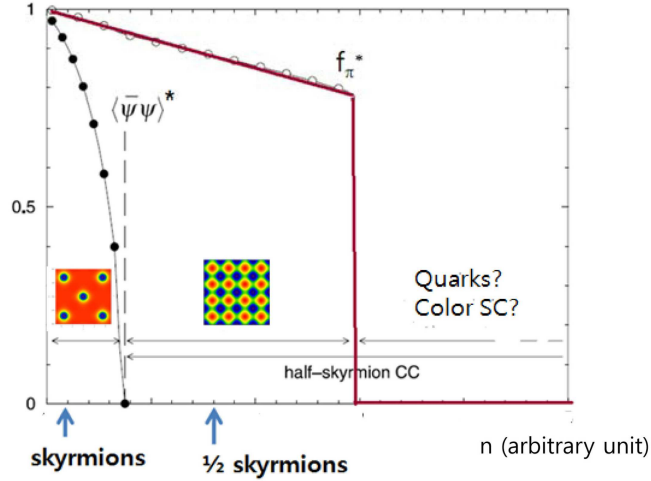}
\caption[0]{Phase diagram near chiral restoration at high density. The relevant densities are in arbitrary units as they depend on the parameters of the Lagrangian which are to be fixed by the physics involved. See Figure 6. }\label{figure5}
\end{figure}

Given the HY HLS Lagrangian, we need to get baryons, and this we can do via topology. The baryons are of course needed for sourcing density. Alternatively one may resort to Weinberg's ``folk theorem" and start with explicit baryon fields and HLS meson fields put in a way fully consistent with chiral symmetry to build dense matter. This approach will be taken up later. What we will do here is to work with these skyrmions generated from the HLS theory. At the moment, there is no continuum treatment using the skyrmions combined with HLS mesons~\footnote{There are attempts to approach dense matter analytically in holographic QCD. In terms of multi-instantons as multi-baryons, one can address such questions as many-body forces at short distance relevant to high density, e.g., Hashimoto et al~\cite{hashimoto}. This approach is being developed and is not yet in a workable form.}. One can, however, put skyrmions on crystal lattice and squeeze the system to generate high density. This matter was discussed by Byung-Yoon Park in his colloquium here and also extensively reviewed elsewhere, so we won't repeat it. What comes out of such treatments is shown in Figure 5\cite{park,PV}. At some density (see later for numbers), there is a phase change from a skyrmion matter -- which is a nucleonic matter -- to a half-skyrmion matter with vanishing quark condensate and non-vanishing pion decay constant. This is a phase in which chiral symmetry is apparently restored (with the vanishing order parameter $\langle\bar{q}q\rangle=0$), but there are hadrons, with presumably parity-doubled and massive baryons and mesons, and the Goldstone pions.~\footnote{It's possible that the 't Hooft anomaly matching is satisfied by the massless pion.} Note that this is a novel phase nobody else has been discussing up to now although some people are arriving at something similar using large $N_c$ arguments, which we think is more of a conjecture than anything else at the moment.

Now you have a dense matter with a particular phase structure which is novel. What happens when you inject a $K^-$ into the system? This sort of experiments, if not done yet,  will be done at the J-PARC and surely at the FAIR/GSI. What happens to an ani-kaon when embedded in the half-skyrmion matter has been worked out. Although the final result appeared after the lecture, let us quote the result here since it is rather striking\cite{kim}. It is given in Figure 6 for the dilaton mass 720 MeV. We do not really know what the dilaton mass we need for our study is. The story of scalar mesons in low-energy strong interaction physics is still very murky and there is no simple understanding. There can be a highly complex interplay between quark degrees of freedom and gluonic degrees of freedom. What is given in Figure 6 is for the dilaton mass that is used in the rather successful mean-field calculation of nuclear matter performed by Chaejun Song\cite{song} using an HLS Lagrangian  in which the vector manifestation, i.e, BR scaling, is implemented in consistency with Landau Fermi liquid theory.

It turns out\cite{kim} that the threshold density for the half-skyrmion phase  $n_{1/2}\approx 1.3n_0$ is more or less independent of the dilaton mass. It is controlled by the parameters of the skyrmion Lagrangian itself. What the dialton mass does is to shift the chiral transition density $n_\chi$: the higher the mass is, the higher $n_\chi$mwill be.
%Figure 6
\begin{figure}[ht]
\centering
\includegraphics[angle=0, width=0.7\textwidth]{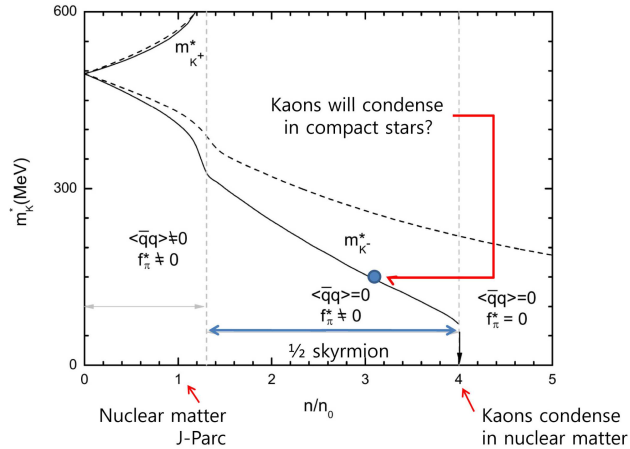}
\caption[0]{Kaon spectrum in the half-skrymion matter for the dilaton mass of 720 MeV.}\label{figure6}
\end{figure}

The important feature to notice is that in a half-skyrmion phase, the kaon mass drops considerably more rapidly than in the usual skyrmion matter, going to zero when the condensate of the soft dilaton -- which coincides with the chiral condensate within the approximations taken -- vanishes. There seems to be, in the half-skyrmion matter, the ``mysterious" agent in Yamazaki's picture that shrinks the matter, a mechanism that is not readily available in ordinary nuclear interactions. Interestingly, kaon condensation can take place even in a symmetric nuclear matter when the scale symmetry breaking tied to chiral symmetry breaking is restored \`a la Freund and Nambu\cite{nambu}, implying that kaon condensation is {\em equivalent} to strange quark matter. If one takes the canonical symmetry energy of \cite{BLRstar}, then kaon condensation will take place at $n
\sim 3n_0$ and things will happen as expected. In fact, the results obtained in this calculation are surprisingly close to what was obtained in the HLS/VM scenario described in \cite{BLRstar,BLR-PRL}. What will happen with Li's ``supersoft" symmetry energy is not clear. It's a nice problem to work out.

\section{Nuclear Matter in ``Double Decimation"}
Going to higher densities in nuclear systems involves a multitude of scales. When there are a multitude of scales involved in many-body processes as in the case of dense-matter physics, one needs to figure out which is the best starting point for a given process. It is clearly senseless to start with physics at nuclear matter density and hope to get, within the same framework, to super-high density where perturbative QCD is applicable,  such as for instance the color-flavor locking phase. There is a good example at high temperature as studied at RHIC. The finding -- which is a total surprise to  many people working in heavy-ion physics -- was that a conformally invariant supersymmetric field theory which has nothing to do with low-temperature world can describe well what's happening at high temperature where they all thought weakly interacting quark-gluon plasma would show up. It appears that whatever symmetries and dynamics that are essential at high temperature for the processes involved there have little to do with what happens at low temperature which can be looked at by lattice QCD. It may be that similar things happen in density.

One thing is certain, and that is that we should be able to go with confidence from zero density up to near the nuclear matter density. Let's see what we can say. Most of the workers in the field start with a Lagrangian determined in matter-free space at zero temperature. Can such a theory go beyond the nuclear matter density and go up to the chiral transition point?

We know that the ordinary nuclear matter we are familiar with is a fixed point of an effective field theory (EFT), namely, Landau Fermi liquid fixed point. Ignoring the possibilities of meson condensation, half-skyrmion (or half-instanton) phase etc, there is next the chiral transition point which in HLS theory is the ``vector manifestation (VM) fixed point." The question is: Can one go from one fixed point to another in a systematic way and if so, how?

It is clear that by just sticking around the matter-free space and insisting on calculating more and more terms -- which is what many nuclear theorists are doing -- one will not get anywhere. One has to find how to proceed in a consistent and efficient way. We don't know how to do this all the way to the critical density for chiral restoration -- and that's why we have been playing around with skyrmion crystals, vector manifestations etc, but we can do it up to, and slightly above, the nuclear matter density. And this is what we will do in this part. How to go from nuclear matter above that density is something that our WCU program will have to figure out.

We will get to nuclear matter by what Gerry Brown and MR called ``double decimation." For this the first thing is Weinberg's ``folk theorem" which was restated this year in \cite{weinbergtheorem}.
\subsection{Weinberg ``folk theorem"}
The theorem goes as follows for the case we are concerned with:

``If we write down the most general possible Lagrangian,
including all terms consistent with chiral symmetry principle,
and calculates matrix elements with the Lagrangian in any
given order of perturbation theory,
the result will simply be the most general S-matrix consistent with
perturbative unitarity, analyticity, cluster decomposition and the assumed symmetry properties"
\\
\\
We will now sketch how this theorem guides us to understand physics of nuclei and nuclear mater.

\subsection{Double decimation}

If we are interested in what's happening in the vicinity of the nuclear matter saturation point, then we can simply start at that point taken as a Fermi liquid fixed point as we will do below. One can write down an effective Lagrangian with correct symmetries, principally chiral symmetry, and calculate weak fluctuations around the Landau Fermi liquid fixed point. This can be phrased using an effective Lagrangian patterned after Walecka's mean field theory. One starts with a suitable HLS Lagrangian with both fermionic and bosonic degrees of freedom that are relevant and work in the mean field. This is shown to be equivalent to doing Fermi liquid theory as we will explain further below. Of course it's not obvious how to go beyond the vicinity of the normal nuclear matter, but the physics just below and just above nuclear matter density should be describable as Walecka's mean-field theory does. One may extrapolate beyond the nuclear matter density in some sense if one endows the parameters of the theory with density dependence. This renders the theory a sort of density functional theory of Kohn-Sham type in field theory language. This is the strategy of BR scaling -- something which has to do with chiral symmetry -- making contact with standard nuclear physics which does not have obvious connection to chiral symmetry. It turns out that this works very well. Bingo for BR scaling!

The question that can be answered is this: Can one go from a Lagrangian defined in the matter-free vacuum to the Fermi liquid fixed point? The answer is yes: It involves the double decimation.

Suppose we have an effective Lagrangian ${\cal L}_{eff}$ which has all relevant degrees of freedom and symmetries required by QCD defined with a cut-off at $\Lambda$. In practice, we are thinking of chiral Lagrangian for which the cut-off is $\Lambda=\Lambda_\chi\approx 4\pi f_\pi\sim 1$ GeV. One usually assumes that in dealing with many-baryon systems, there is a Fermi sea defined by the Fermi momentum $k_F$. In principle, the Fermi sea should come out of the theory, as most likely a quantum critical phenomenon. The only discussion on this matter in connection with nuclei and nuclear matter as far as we know was made by Lynn in 1993~\cite{lynn} where he suggested that a drop of nuclear matter, i.e., nucleus, could arise as a non-topological soliton, a chiral liquid from the chiral Lagrangian with which the effective potential is calculated in chiral perturbation theory. In practice, one cannot do this to all orders since an effective theory defined with a cut-off would require an infinite number of parameters for an all-order calculation. Even when drastically truncated, the theory could not quite arrive at a form that resembles the requisite Fermi liquid, and no further work along the line initiated by Lynn has been done since.
\begin{figure}[ht]
\centering
\includegraphics[angle=0, width=0.9\textwidth]{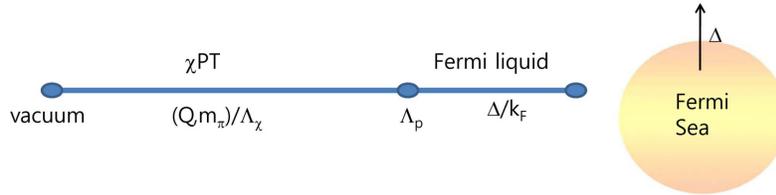}
\vskip -0.5cm
\caption[0]{Multiple scales and expansion parameters. Here $Q$ is the typical momentum/energy scale probed, $m_\pi$ is the pion mass due to the  explicit breaking of chiral symmetry by the quark mass. Note that up to $\Lambda_p$, the Fermi momentum appears as a ``small" expansion parameter but near the Fermi surface, it appears as a ``large" $N$ in $1/N$ expansion in the Fermi-liquid fixed point Lagrangian. }\label{figure7}
\end{figure}

A program that has made a big progress is that of Gerry Brown and his collaborators. They assume that there is a Fermi sea and derive nuclear matter in two steps, in what Gerry and MR called ``double decimation." This is indicated in Figure 7 with the pertinent expansion parameters. The first step is from zero energy scale ($\Lambda=0$) roughly to the scale up to which nucleon-nucleon (NN) elastic scattering is measured and the second is from the free-space interactions to the Fermi sea scale with many-body interactions. Here one is doing an effective field theory, making the expansion in terms of some small expansion parameters. Now there is an intricacy in the way the expansion is to be done in the two steps. The first resorts typically to chiral perturbation theory where the expansion parameters are $Q/\Lambda$ and $m_\pi/\Lambda$ with  $Q$ being the small probe momentum, $m_\pi$ small chiral symmetry breaking and $\Lambda$ the chiral scale $\sim 4\pi f_\pi$. In this step, the Fermi momentum $k_F$ is considered as ${\cal O}(Q)$. However in the second step where one will ultimately have to perturb around the Fermi surface, $k_F$ comes as $1/k_F$, counted as ${\cal O} (1/N)$ where $N$ is to be considered as large. This is indicated in Figure 7.

Let us describe these two steps in some detail.
\subsection{$V_{lowk}$}
Nuclear potential is not a measurable quantity; it is however a convenient intermediate tool to get to measurable quantities. It's a nice way of translating the large amount of NN scattering data given in terms of phase shifts, accumulated by experimentalists over the years,  into physical observables in many-body systems. Assuming that two-body interactions are dominant in many-nucleon systems, the potential extracted from scattering data plays a key role in calculating the properties of nuclei and nuclear matter. We will come to n-body forces for $n > 2$ below. Here we will continue ignoring $n > 2$.

There are in the literature many such ``realistic potentials" constructed from the given experimental phase-shift data, more or less guided by certain theoretical framework, such as meson exchanges, dispersion relations etc. The examples are Bonn A, CD-Bonn, Paris, Nijmegen I and II,  Argonne v18, chiral perturbation, Idaho A etc. They are all fine-tuned with a certain number of free parameters~\footnote{The potentials derived from chiral perturbation theory at low orders have fewer parameters, but going to higher order would require an unmanageable number of parameters.} to fit the elastic scattering data, typically up to the lab momentum $p_{max}\sim 300-350$ MeV. The fit is achieved with $\chi^2$/dof $\approx 1$, some better than others. One thing common among all these ``realistic potentials" is that the long-range part has to be consistent with one-pion exchange. However the short-range parts can be drastically different from one potential to another. If one were to naively use these potentials in perturbation expansion to compute the properties of nuclei and nuclear matter, the short-distance components of the potential would come in very importantly, which account for the tremendous number of papers written on now to do many-body calculations in nuclear physics.

An adroit use of  $V_{lowk}$ resolves this conundrum. How this comes about is the story of Wilsonian renormalization group (RG) technique.

An effective theory is defined with a cutoff $\Lambda$ above which the theory is not valid. This is clearly defined in chiral perturbation approaches but with other phenomenological potentials, this is not the case. To be as precise as possible, we will put the arguments in the framework of effective field theory.

Effective field theories dealing with chiral dynamics have the cutoff at the chiral scale $\sim 4\pi f_\pi$. Suppose we have a two-body potential $V_{NN}$ determined by fitting elastic scattering data up to the momentum $p_{max}$. The NN scattering amplitude ${\cal A}$ is given by the infinite sum
\begin{eqnarray}
\la f|\A|i\ra = \la f|V_{NN}|i\ra + \sum^{\infty}_{n=0} \frac{\la f|V_{NN}|n\ra \la n|V_{NN}|i\ra}{E_{i}-E_{n}}+\cdots\label{t-matrix}
\end{eqnarray}
where the ellipsis stands for the infinite higher order terms.
Here the intermediate sum is written to go up to infinity but it should actually be limited to the cutoff scale, in principle well-defined in EFT, but in phenomenological potentials, the cutoff is done with form factors etc. which cannot be given a precise meaning. Now the maximum probe momentum is $p_{max}$ up to which precise data are available, so for observables probed at $p\ll p_{max}$, it makes sense to lower the cutoff from $\Lambda_\chi$ (or $\infty$) to $\Lambda_{NN}\sim p_{max}$. One can rewrite (\ref{t-matrix}), defining an effective low-momentum potential $V_{lowk}$ which results from integrating out high energy mode, as~\footnote{This is the first decimation in the the double decimation we are doing.}
\be
\la f|\A|i\ra = \la f|V_{lowk}|i\ra + \sum^{\Lambda_{NN}}_{n=0} \frac{\la f|V_{lowk}|n\ra \la n|V_{lowk}|i\ra}{E_{i}-E_{n}}+\cdots
\ee
Note that $\Lambda_{NN}$ is arbitrary, so physics should not depend on where $\Lambda_{NN}$ is put. This gives the renormalization group equation (RGE)
\be
\frac{d\la f|\A|i\ra}{d\Lambda_{NN}}=0.\label{rge}
\ee
How the effective potential runs with the scale $\Lambda_{NN}$ is given by
\be
\frac{d}{d\Lambda_{NN}} V_{lowk}=\beta([V_{lowk}], \Lambda_{NN}).
\ee
The $\beta$ function is then calculated order by order in an EFT scheme, e.g., chiral perturbation. Note that when the $\beta$ function vanishes, one has a fixed point. We will have this fixed point in the second decimation.

This is all fine in principle. But in practice, one is limited by certain calculational techniques; the $\beta$  function cannot be calculated to all orders, so it's not simple to satisfy Eq.~(\ref{rge}). The strategy is then to search for $\Lambda_{NN}$ where the RGE is best satisfied. One finds that this comes out well for $\Lambda_{NN}\approx 2.1$ fm$^{-1} \approx 3 m_\pi$.

Given a potential $V_{NN}$, calculating the corresponding $V_{lowk}$ cannot always be done analytically. In all cases except in special approximations, it can only be done numerically. It is instructive to construct a schematic model for the potential of different forms that can be handled analytically which can be fit to experimental data in a particular channel. It turns out that this can be done for the $^1S_0$ phase shift. In  \cite{bogner}, two separable potentials called I and II were constructed to fit  via Eq.~(\ref{t-matrix}) the $^1S_0$ phase shift. While the $V_{NN}$'s so obtained give quite different diagonal matrix elements as shown in Fig.~8 (left panel), the $V_{lowk}$'s resulting from the decimation give identical matrix elements as shown in Fig.~8 (right panel). Also shown there are the results obtained with the realistic potentials, which all converge on the same curve if one takes $\Lambda_{NN}=2.1$ fm$^{-1}$.\footnote{For different values of $\Lambda_{NN}$, not all potentials agree in $V_{lowk}$ in various channels. It's not clear why $\Lambda_{NN}=2.1$ fm$^{-1}$ gives a better convergence than other values that are somewhat higher or somewhat lower. It could have something to do with various other degrees of freedom starting to come into the picture.}
\begin{figure}[ht]
\centering
\includegraphics[angle=0, width=0.8\textwidth]{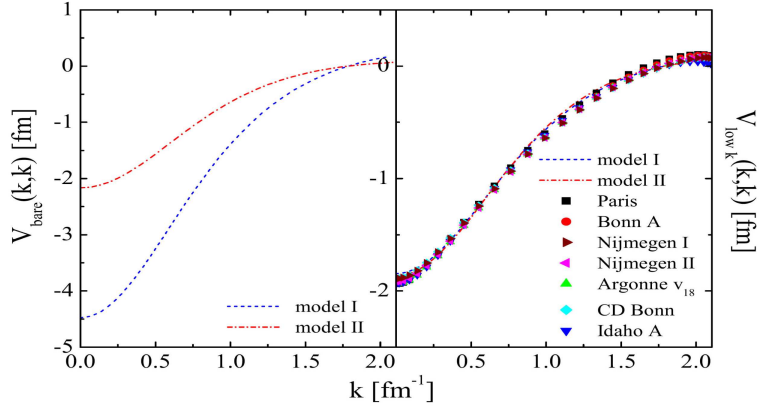}
\caption[0]{The diagonal momentum-space matrix elements for separable potentials I and II.}\label{figure8}
\end{figure}
\subsection{Landau Fermi liquid}
Given the $V_{lowk}$ determined in  matter-free space from NN scattering, how would one go about describing many-nucleon systems, i.e., nuclei and nuclear matter? In medium, $V_{lowk}$ will get modified by change in the vacuum structure, i.e., by BR scaling, and/or renormalized by many-body processes. The latter could be calculated in a systematic many-body theory. But there also can be intrinsic many-body forces, not accounted for in many-body scatterings. For instance in chiral perturbation theory, there are three-body forces that appear at $\O ((Q/\Lambda_\chi)^3)$, Figure 9. We will not have anything to say how three-body and higher-body forces should figure in the RG procedure. There is very little done in that  direction. Let us proceed with the two-body $V_{lowk}$ alone.

Consider describing many-body systems in which nucleons are interacting via $V_{lowk}$ in a Fermi sea. We don't know how the Fermi surface arises. We will simply assume that there is a Fermi sea formed somehow in the theory. We are interested in computing the Bethe-Salpeter equation for fully irreducible vertex $\Gamma$ in terms of particle-hole irreducible vertex $\tilde{\Gamma}$ given in Figure 10. In calculating the integral equation, the momentum cutoff is set to $\Lambda_{NN}=2.1$ fm$^{-1}$. The Landau Fermi liquid theory is given by the vertices in the so-called ``$k$ limit" and ``$\omega$ limit": $\frac{|\vec{K}|}{w}\rightarrow 0$ and $\frac{w}{|\vec{K}|}\rightarrow 0$ where $K=(\omega, \vec{K})$~\cite{abrikosov},
\begin{eqnarray}
\Gamma^{w}(k_{1},k_{2})=\lim_{w\rightarrow 0}\lim_{|\vec{K}| \rightarrow 0} \Gamma(k_{1},k_{2};K)\ ,
 \nonumber \\
\Gamma^{k}(k_{1},k_{2})=\lim_{|\vec{K}|\rightarrow 0}\lim_{w \rightarrow 0} \Gamma(k_{1},k_{2};K).
\nonumber
\end{eqnarray}
\begin{figure}[ht]
\centering
\includegraphics[angle=0, width=0.8\textwidth]{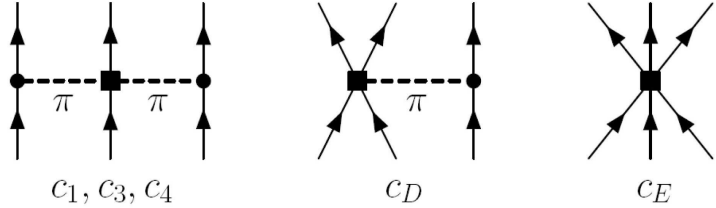}
\caption[0]{Three-body forces that contribute in chiral perturbation theory at $\O ((Q/\Lambda_\chi)^3)$. The constants $C_{1,3.4}$ are determined from $\pi$N scattering and $C_D$ can be determined via PCAC from the axial current matrix element of two-body systems. The constant $C_E$ is the only undetermined parameter in free space.}\label{figure9}
\end{figure}

The forward scattering amplitude $\Gamma^k$ is given by
\be
\Gamma^k_{\alpha\beta,\gamma\delta} (k_1,k_2)=\Gamma^\omega_{\alpha\beta,\gamma\delta} - \frac{1}{16\pi}N_0 Z^2\int d\Omega_{q} \Gamma^{\omega}_{\alpha \epsilon \delta \eta}(k_{1},q)
\Gamma^{k}_{\eta \beta, \epsilon \delta }(q, k_{2}).
\ee
What is known as ``Landau $f$ function" is
\be
f(k_1,k_2)\equiv Z^2 \Gamma^\omega (k_1,k_2)
\ee
and ``Landau $a$ function" -- which is forward scattering amplitude
\be
a(k_1,k_2)\equiv Z^2 \Gamma^k (k_1,k_2).
\ee
In nuclear physics, both spin and isospin figure, so the Landau $f$ function has the components (in condensed matter, the isospin terms are absent)
\be
f(k_1,k_2)=\frac{1}{N_0}\left[F(k_1,k_2) + F^\prime (k_1,k_2)\tau_1\cdot\tau_2 + G(k_1,k_2)\sigma_1\cdot\sigma_2 + G^\prime (k_1,k_2)\tau_1\cdot\tau_2 \sigma_1\cdot\sigma_2\right].\nonumber
\ee
There are also tensor terms but we will not write them down.

Given these Landau functions, a small change in the single particle distribution function $\delta n(k)$ induces the change in energy of the form
\be
\delta E=\sum_k \epsilon^{(0)}_k\delta n(k)+\frac{1}{2V}\sum_{k_1,k_2} f(k_1,k_2)\delta n(k_1)\delta n(k_2) +\O ((\delta n(k))^3)
\ee
From this one can calculate all the bulk properties of nuclear matter. For instance, some relevant physical observables are
\be
\frac{m^*}{m}&=&1+\frac{F_1}{3}, \\
\beta &=& \frac{\hbar^2 k_F^2}{6m^*} (1+F_0^\prime),\\
K &=& \frac{3\hbar^2 k_F^2}{m^*} (1+F_0),\\
\delta g_l&=& \frac 16 \frac{F_1^\prime -F_1}{1+F_1/3}
\ee
where $F(k,k^{\prime}) = \sum_{l} F_{l}P_{l}(\cos\theta)$. They correspond, respectively, to the effective mass of the quasiparticle, symmetry energy, compression modulus and anomalous orbital gyromagnetic ratio.

\begin{figure}[ht]
\centering
\includegraphics[angle=0, width=0.7\textwidth]{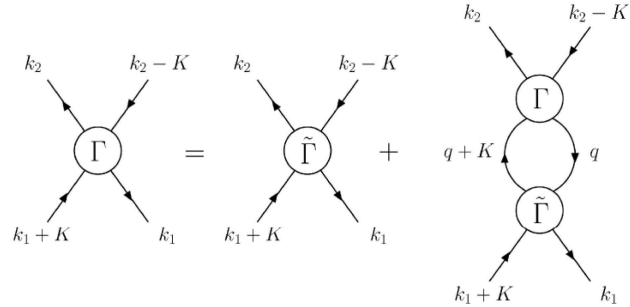}
\caption[0]{Bethe-Salpeter equation}\label{figure10}
\end{figure}

These Landau $f$ functions have been computed by summing Feynman diagrams using the technique of induced interaction with $V_{lowk}$ in \cite{hhkb}. What one finds is quite interesting and important.

When calculated with the $V_{lowk}$ determined in matter-free space (elastic scattering) and with no 3-body forces, the results (given in the first column of (\ref{results})) are not quite satisfactory. One notes that the discrepancy in $\delta g_l$ is particularly serious.
\subsection{Enter BR scaling and/or three-body forces}
There are two possible ways to improve these results~\cite{hhkb}. One is to incorporate BR scaling in $V_{lowk} $ which is to endow it with density dependence, i.e., $V_{lowk} (n)$. The way to do this is to consider one-boson exchange potentials -- such as Nijmegen potential I and II and CD-Bonn potential -- and endow BR scaling to the mass of the exchanged particles. In doing this, one assums that there is no scaling in the coupling constants. There is no good reason why this is OK but this turns out to be fine up to $n\approx n_0$. However as was discussed in one of the previous lectures, hidden local symmetry with vector manifestation \`a la Harada and Yamawaki predicts that at least the vector meson mass $m_V$ and the coupling $g_V$ run in density proportionally to each other as $g_V/m_V\sim const.$ for density $n > n_0$. Scalar meson masses could in principle scale differently. However the mass of the dilaton scalar $\chi_s$ connected with the trace anomaly of QCD does run like vector mesons in the skyrmion crystal matter. This property remains to be confirmed more rigorously. With the BR-scaled Landau $f$ functions, the results (second column of (\ref{results})) are considerably improved.
\begin{eqnarray}
\begin{array}{cccc}
\textnormal{At n} = n_{0} & & &  \\
 & \textnormal{No BR scaling}  & \textnormal{With BR scaling} & \textnormal{Exp.}\\
\beta &  \simeq 18-19\ {\rm MeV} & 20 -25\ {\rm MeV} & 25-26\ {\rm MeV} \\
\frac{m^{\star}}{m} & \simeq 0.9  & 0.7 & 0.6-0.7\\
K & \simeq 100-136\ {\rm MeV}  & 140-210\ {\rm MeV} & 200-300\ {\rm MeV} \\
\delta g_{l} & \simeq 0.7  & 0.2-0.3 & 0.23 \pm 0.07\label{results}
\end{array}
\end{eqnarray}
    The improvement in $\delta g_{l}$ is particularly noteworthy. We will return to this quantity below in connection with doing Landau fixed point theory in one-go rather than in double decimation so far discussed.

There is a possibility that one can obtain similar improvements by introducing three-body and higher-body forces in lieu of BR scaling.~\footnote{N-body forces with N $>3$ may also contribute but we will not consider them here. They are expected to be suppressed by chiral counting but they could also be suppressed  dynamically. We suggest that both BR scaling and 3-body forces should be included together rather than one or the other.} To see how 3-body forces enter in this EFT/RG formulation, it would be necessary to develop a $V_{lowk}$ for three-body systems. This is because short-range interactions which would be badly behaving need to be tamed by an RG procedure as in the case of 2-body force. It's not at all obvious how this can be done given the paucity of both experimental and theoretical inputs. We know from the C14 dating problem that in certain channels,  BR scaling and three-body forces do a similar thing and can explain more or less similarly the long life-time of the C14. This suggests strongly that they are not independent of each oteher. In fact, Song in his thesis showed that BR scaling implemented in two-body EFT yields correct thermodynamic consistency conditions because what is known as ``rearrangement term" involving effective many-body forces is implicitly included in the density dependence of the mass and coupling constant.

\begin{figure}[h]
\centering
\includegraphics[angle=0, width=0.5\textwidth]{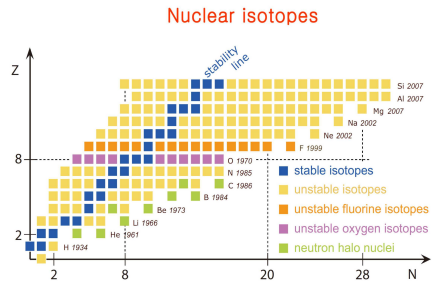}
\caption[0]{Nuclear isotopes to be charted by the RIB machines in the world including the ``KoRIA."}\label{figure11}
\end{figure}
\begin{figure}[h]
\centering
\includegraphics[angle=0, width=0.6\textwidth]{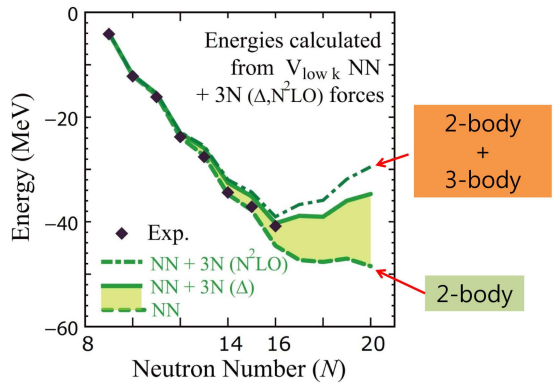}
\caption[0]{Ground state energies of neutron-rich oxygen isotopes. Note the important role of three-body forces for neutron number greater than 16.~\cite{otsuka}}\label{figure12}
\end{figure}

Let us mention a recent interesting observation that deserves further study. It may have a direct relevance to the KoRIA project~\footnote{This matter was not addressed in detail in this lecture but we think it's worth being added here.}. It has to do with the neutron drip line, in particular in the oxygen series~\cite{otsuka}. Shown in Figure \ref{figure11} is the chart of nuclear isotopes measured and understood as well as those to be measured in the future RIB machines. Understanding neutron-rich nuclei will help understand the nuclear symmetry energy that is indispensable for the physics of neutron stars discussed above. The paper \cite{otsuka} addresses the neutron drip line in oxygen isotopes and argues that the three-body forces of Fig.~9 are needed to explain the present status with the drip line. See Figure \ref{figure12}. One notes that it is the long-range part of the three-body forces that does most of the job, while the shortest-ranged one with the coefficient $C_E$ reinforces the tendency. This however seems different from the case of C14 dating where the contact term figures more importantly. It will be interesting to see what the cause of this difference is. Here is a conjecture: that incorporating BR scaling {\em without three-body forces} could do the same job in the neutron drip line phenomenon.
\subsection{BR scaling in tensor forces}
We know that tensor forces are very important in nuclear physics for various observables in nuclear processes. They are considered to be important also beyond the nuclear matter density for compact stars, namely in the EOS, e.g., the symmetry energy. We know how the tensor forces would scale up to nuclear matter density but how the $\rho$ tensor force scales above the nuclear matter density, in particular, from the density at which the half-skyrmion phase (equivalently the ``hadronic freedom (HF)") sets in is not known. Note that this is the density regime where the ``supersoft" symmetry energy turns over.

Let us see what the ``double decimation" strategy says about the scaling behavior. The tensor forces coming from the pion and $\rho$ exchanges are given in the form
\be
V^T_{\rho} &=& \frac{f^2_{N\rho}}{4\pi} m_{\rho} (\tau_1 \cdot \tau_2) (-S_{12})\left[\frac{1}{(m_{\rho}r)^3} + \frac{1}{(m_{\rho}r)^2} + \frac{1}{3m_{\rho}r} \right] \exp^{-m_{\rho} r}, \label{vtrho} \\
V^T_{\pi} &=& \frac{f^2_{N\pi}}{4\pi} m_{\pi} (\tau_1 \cdot \tau_2) S_{12}\left[\frac{1}{(m_{\pi}r)^3} + \frac{1}{(m_{\pi}r)^2} + \frac{1}{3m_{\pi}r} \right] \exp^{-m_{\pi} r}. \label{vtpi}
\ee
In dense medium, the pion tensor is assumed to remain more or less unchanged but the $\rho$ tensor undergoes the BR scaling. This is of course strictly correct in the chiral limit but can only be an approximation when the pion mass is taken into account. We will assume that it's OK. In Holt et al.\cite{holtetal} where the long-standing C14 dating is explained and in Xu and Li\cite{xu-li} where the ``suspersoft" symmetry energy is shown to be understandable in terms of BR scaling in the $\rho$ tensor force, the $\rho$ mass is taken to scale in (\ref{vtrho}), but the coupling constant $f_{\rho N}$ is taken non-scaling.

Let us see what the current version of BR scaling says. In HLS-type theory, one has
\be
f_{N\rho}=(1+\kappa_\rho)\frac{m_\rho g_{\rho NN}}{2m_N},
\ee
Here $\kappa_\rho g_{\rho NN}$ is the anomalous tensor $\rho NN$ coupling and $g_{\rho NN}$ is related to the hidden gauge coupling, so we will simply denote it as $g$. Thus one is interested in the scaling
\be
f^*_{N\rho}/f_{N\rho}\approx \frac{m^*_\rho}{m_\rho}\frac{m_N}{m^*_N}\frac{g^*}{g}.\label{scaling}
\ee
Now both empirically and theoretically, the nucleon mass scales up to $n_0$ as
\be
m^*_N/m_N\approx f^*_\pi/f_\pi.
\ee
and BR scaling says, also up to $n_0$,
\be
m^*_\rho/m_\rho\approx f^*_\pi/f_\pi.\label{N-scaling}
\ee
Thus up to $n_0$,
\be
f^*_{N\rho}/f_{N\rho}\approx \frac{g^*}{g}\approx 1\label{scalingg}
\ee
where the last equality follows from ``double decimation." In \cite{holtetal}, this scaling is fine since there density only up to $n_0$ is probed. However in \cite{xu-li}, this scaling is taken beyond $n_0$ and we will now argue this is not consistent with the modern BR scaling.

First of all, once one reaches the density corresponding to the half-skyrmion density (or HF density), the second equality in (\ref{scalingg}) does not hold. In fact, HLS/VM says that it should scale like $\la\bar{q}q\ra^\star/\la\bar{q}q\ra$ which scales as $m^*_\rho/m_\rho\equiv \Phi$. Next, one has to have an idea as to how the nucleon mass scales beyond $n_0$. In the old version of BR scaling, it was taken to scale like $\Phi$ which means that in the chiral limit it should vanish. But this is at variance with the half-skyrmion picture which goes as follows. In the skyrmion matter description, the single-nucleon mass in the matter must go like
\be
m_N^\star\sim f_\pi^\star \kappa
\ee
where $\kappa$ is an ${\O}(\sqrt{N_c})$ term consisting in general of a scale-invariant piece (e.g., the $1/e$ coefficient of the Skyrme term in the Skyrme Lagranigan) and scale-symmetry-breaking (SSB) corrections. If we ignore the SSB corrections, then the nucleon mass scales as (\ref{N-scaling}) even in the half-skyrmion phase up to the chiral transition point. Now we know that in the half-skyrmion phase, $f_\pi^\star$ remains non-zero although the quark condensate goes to zero and furthermore it scales little if any. So we may simply let the scaling stop at $n_0$ and take it a constant,
\be
f^*_{N\rho}/f_{N\rho}\approx \frac{m_N}{m_N^\star (n_0)}\frac{m^*_\rho}{m_\rho}\frac{g^*}{g}\approx (1/b)\Phi^2\label{scalinggg}
\ee
with
\be \Phi=\frac{m_\rho^\star}{m_\rho}\approx \frac{\la\bar{q}q\ra^\star}{\la\bar{q}q\ra}.
\ee
Taking the empirical value, we will have
 \be
 b=m_N^\star (n_0)/m_N\approx 0.7 \ \ {\rm for}\ \ n>n_{1/2}.
 \ee
This means that once one is in the half-skyrmion phase, the $\rho$ tensor is suppressed by the VM property even though the $\rho$ mass drops. In fact, a rough estimate shows that the $\rho$ tensor will be almost completely suppressed by $n=3n_0$, just where the ``supersoft" symmetry energy drops to negative value. It appears certain that this will change the Bao-An Li's scenario on the non-Newtonian gravity.
\subsection{Parity-doubling model for the nucleon}
The observation that the nucleon mass $\propto f_\pi^\star$ in the half-skyrmion phase stays non-vanishing up to the chiral transition density suggests that the effective chiral Lagrangian with explicit baryon fields could have a non-zero chiral invariant mass $m_0$ as in the parity doubler model of DeTar and Kunihiro\cite{detar-kunihiro}. Work on this picture is going on at present, and there is nothing specific to report.

The general picture, however, is not difficult to figure out.

For consistency with what we have been doing in this lecture, we need to transform DeTar-Kuniharu's linear sigma model to non-linear sigma model which is gauge equivalent to hidden local symmetry theory that we are working with. To do that, define
\be
\Sigma=\frac 1{f_\pi}(\sigma+i\vec{\tau}\cdot\vec{\pi})=\xi^2
\label{D1}
\ee
in the unitary gauge with $\xi_R^{-1}=\xi_L=\xi$.  The baryon doubler field $\Psi$
in the linear representation relates to the baryon doubler field ${\bf Q}$ in the non-linear
representation through $\Psi=\xi_5^{-1}{\bf Q}$ with
\be
\Sigma_5=\frac 1{f_\pi}(\sigma {\bf 1}_2+i\vec{\tau}\cdot\vec{\pi}\Gamma_5)=\xi_5^2
\label{D2}
\ee
where $\Gamma_5=\rho_3\otimes \gamma_5$.  The $\rho$'s are Pauli matrices acting on the baryon
doubler (reserving $\tau$ for the Pauli matrices for isospin) and the $\gamma$'s act on the usual Dirac spinor.  Thus the $SU(2)$ non-linear version of the model of \cite{detar-kunihiro}  without explicit chiral symmetry breaking is
\be
{\cal L}_2  =&&\overline{\bf Q}i\rlap/\partial {\bf Q}-g_1f_\pi\overline{\bf Q}{\bf Q}
+g_2f_\pi\overline{\bf Q}\rho_3{\bf Q}\nonumber\\
&&-im_0\overline{\bf Q}\rho_2\gamma_5{\bf Q}+i\overline{\bf Q}\gamma^\mu{\bf V}_\mu{\bf Q}
+g_A\overline{\bf Q}\Gamma_5\gamma^\mu{\bf A}_\mu{\bf Q}
\label{D3}
\ee
where the vector and axial currents are defined as
\be
{\bf V}_\mu=&&\frac 12 (\xi^\dagger\partial_\mu\xi+\xi\partial_\mu\xi^\dagger)\nonumber\\
{\bf A}_\mu=&&\frac 12 (\xi^\dagger\partial_\mu\xi-\xi\partial_\mu\xi^\dagger).
\label{D4}
\ee
In (\ref{D3}), the $g_1$ and $g_2$ terms are the dynamically generated baryon masses and $m_0$ is the chiral invariant mass. One can think of the dynamical mass vanishing as the quark condensate vanishes, so in the half-skyrmion phase the baryon mass will be given by $m_0$. There are attempts in the literature to verify whether this model is empirically viable\cite{rischke}. So far one cannot say. As of today, the reasonable value for $m_0$ is $m_0\sim 0.5-0.7$ GeV\cite{rischke,pisarski}. Now when applied to dense matter in the sense of mean field, if one assumes that the dynamically generated mass goes to zero at the density where the half-skyrmion phase sets in,  this would give the constant $b$ in (\ref{scalinggg}),
\be
b\sim 0.5 - 0.7 \ \ {\rm for}\ \ n>n_{1/2}.
\ee
This is tantalizingly close to what's gotten in the half-skyrmion crystal calculation.

Although in this lecture we do not go near the chiral transition density for which we are not ready yet, there is a very simple and beautiful consequence of the scenario given above (i.e., the half-skyrmion or parity-doubled baryon matter) as one approaches chiral restoration. It is that the massive baryons will not play any role in the thermodynamics of the system. If the nucleon mass remains relatively large, say, $\sim 700$ MeV, compared with the meson masses that are dropping \`a la BR scaling, once the density exceeds $n_{1/2}$, then the baryons can be integrated out, which will thereby endow the parameters of the HLS Lagrangian with density dependence. In this case, the properties of the mesons in dense medium approaching the chiral transition density $n_\chi$ will satisfy the same formulae as the ones in the temperature case except that now $T$ can be replaced by density $n$. This means that the same VM property should hold in density as in the temperature case. This feature will play an important role in going toward compact-star regime.

\section{Landau Fermi-Liquid Fixed-Point Theory}
If one accepts that nuclear matter is at the Fermi-liquid fixed point as explained, e.g., by Shankar~\cite{shankar}, one can get to the answer simply via mean-field theory of EFT. To do this, one fixes the Fermi momentum $k_{F}$, which means fixing the density at $n_{0}$. One can then write an effective Lagrangian valid near the Fermi surface as we mentioned in connection with Figure 7. Consider having integrated out all degrees of freedom other than nucleons and write the EFT action as an expansion in fermion field
\begin{eqnarray}
\mathcal{S} = \int \frac{dk d\omega d\Omega}{(2 \pi)^{4}} \bar{\psi} \left( iw - v^{\star} k \right)\psi
+ \int U\bar{\psi} \bar{\psi} \psi\psi + \O ((\bar{\psi}\psi)^3)\label{Landau}
\end{eqnarray}
with $v^{\star} = \frac{k_{F}}{m^{\star}}$. Here the expansion is with respect to the Fermi surface at $k_F$ which is taken to be large, so that $\Delta/k_F\sim 1/N\rightarrow 0$. Take the kinetic energy term -- bilinear in fermion field -- with $v^*$ fixed to be scale invariant; then a careful analysis~\cite{shankar} shows that the four-Fermi interaction is marginal only when the 4-point vertex $U$ is a constant $U_0$ in momentum space. All momentum-dependent terms~\footnote{They contain non-forward scattering amplitude $F(\phi\neq 0)$ in the Landau function.} are irrelevant in the sense of RG, so scale to zero by RG flow. The constant $U_0$ is -- modulo an overall constant -- just the Landau $f$ function we used above. The action (\ref{Landau}) is then an EFT for  Fermi liquid, with the fixed point describing the nuclear matter saturation. All we discussed above will then follow from this Lagrangian.

Now here is the punch line. The fixed point theory (\ref{Landau}) is {\em equivalent} to a mean-field theory \`a la Walecka of HLS Lagrangian in which fermions are introduced as the relevant degrees of freedom~\footnote{Recall this follows from Weinberg's f theorem as MR discussed in other lectures. The basis of this argument is the work of Matsui~\cite{matsui} who discussed the equivalence between Landau Fermi-liquid theory and Walecka's mean field theory.}. First gauge-fix the HLS Lagrangian to unitary gauge and write Walecka's mean-field Lagrangian as
\begin{eqnarray}
\mathcal{L} =\bar{\Psi} \left[ \partial_{\mu} \left(i \partial^{\mu} - g \omega^{\mu} \right) \right] \Psi
+ \frac{1}{2}\left( \partial_{\mu} \phi \partial^{\mu} \phi - m_{s} \phi^{2}\right) - \frac{1}{4}F_{\mu\nu}F^{\mu\nu}
+\frac{1}{2}m_{\omega}^{2} \omega_{\mu} \omega^{\mu} +\cdots\label{walecka}
\end{eqnarray}
 We have dropped all other fields (such as pion, $\rho$, etc) that can enter in general, i.e., in asymmetric nuclear systems. The scalar field $\phi$ in this Lagrangian is not the usual scalar denoted as $\sigma$ in the linear sigma model but a scalar $\chi_s$ that is connected to the trace anomaly of QCD (explained by Hyun Kyu Lee and MR recently~\cite{LR}). As it stands, Eq.~(\ref{walecka}) -- without filling in the ellipsis -- does not give a good description of nuclear matter. In fact, it gives too stiff an EOS with $K\approx 500$ MeV. The remedy to this defect is to incorporate BR scaling as we will do below or incorporate many-body forces generated by higher dimension operators of the type
\begin{eqnarray}
\delta \mathcal{L} = \frac{1}{3} g_{2}\phi^{3} +\frac{1}{4} g_{3} \phi^{4}\label{higher}
\end{eqnarray}
where $g_{2,3}$ are unknown parameters. By adjusting the free parameters of this Lagrangian, one can obtain, in mean field, an impressive fit of nuclear structure from light nuclei to nuclear matter. Just to have an idea of how well it works, take a look at the reference~\cite{lala}. We would conjecture that putting BR scaling suitably in (\ref{walecka}) without (\ref{higher}) would do equally well. Although this has not been worked out fully, the thesis of Song lends a partial support to this conjecture.

Let us finish this lecture by showing what the mean-field approach with a Lagrangian of the type (\ref{walecka}) endowed with BR scaling does for the anomalous gyromagnetic ratio considered above. This matter is discussed in many places, so we won't go into details (which can be found in \cite{songetal}). Putting in the pion field into (\ref{walecka}) in a chirally invariant way, one gets for the proton
\be
\delta g_l^p=\frac{4}{9}\left[\Phi (n_0)^{-1}-1 - ``{\rm pion term}" (n_0)\right]
\ee
where ``pion term" is given as a function of density by chiral symmetry, accurately known at low density, and $\Phi$ is the BR scaling given at $n=n_0$. Using $\Phi(n_0)\approx \frac{f_\pi^*(n_0)}{f_\pi}\approx 0.78$  from pionic atoms and ``pion term"$(n_0)\approx -0.23$, one finds that $\delta g_l^p\approx 0.23$. This was considered as the first indication that BR scaling is operative in nuclei.
\section{Three-Body Repulsion at High Density}
This part was only briefly touched on in the lecture but let us add it. We think it will be one of the major issues to resolve if one wants to get a realistic EOS for neutron stars.

The repulsion at short distance which presumably figures importantly at high density -- and which should be present as is clear from the contact three-body term in Figure 9 -- must clearly be tamed as was done in the case of 2-body repulsion. At present, this has not been done in RG formalisms: There is no 3-body $V_{lowk}$. There are indications from HLS/VM in dense medium and also holographic QCD at short distances that there should be suppression of such repulsion. In HLS theory in QCD as well as in hQCD, the contact three-body force is expected to be given by the exchange of vector mesons. At high density, the gauge coupling $g$ will go to zero in HLS/VM, so there will be an intrinsic suppression of the repulsion by the vector manifestation. In hQCD, as shown by Hashimoto~\cite{hashimoto}, N-body forces are suppressed at short distance -- and in consequence, at high density -- as $\O (N_c/\lambda^{N-1})$ where $\lambda$ is the 't Hooft constant $\lambda=g_{YM}^2 N_c$. Since in hadrons, $N_c/\lambda\sim 1/10$,  many-body forces will be strongly suppressed. This hQCD argument is indicative of an exact many-body solution as it involves a multi-instanton solution that is analytic and exact at short distances in the model when the space becomes flat, so in some sense indicative of what could come out in realistic theories with a generic 5D YM structure (which may be shared by QCD). In going to higher density toward neutron stars and ultimately to $n_\chi$, this is one of the most important issues we will have to resolve.

\subsection*{Acknowledgments}
This work was supported by the WCU project of Korean Ministry of Education, Science and Technology (R33-2008-000-10087-0).

%\section*{References}

\end{document}